\documentclass[aps,a4paper,showpacs,showkeys]{article}
\usepackage[twoside,top=2cm,bottom=2cm,left=2cm,right=2cm]{geometry}
\usepackage[utf8]{inputenc}	
\usepackage{authblk} 
\usepackage{epsfig}
\usepackage{amsmath}
\usepackage{amsfonts}
\usepackage{amssymb}
\usepackage{graphicx}
\usepackage{colordvi}
\usepackage{widetext}
\usepackage[title]{appendix}
\date{}
\begin{document}
	\author{Sergio De Filippo$^{(1)}$\footnote{%
			e-mail address: \textit{sdefilippo@sa.infn.it}}, Filippo Maimone$^{(2)}$\footnote{%
			e-mail address: \textit{filippo.maimone@gmail.com}}, Adele Naddeo$^{(3)}$\footnote{%
			e-mail address: \textit{anaddeo@na.infn.it}}, Giovanni Scelza$^{(2)}$\footnote{%
			e-mail address: \textit{lucasce73@gmail.com}}}
	\affil{\small{\textit{$^{(1)}$Dipartimento di Fisica ``E.R. Caianiello",
				Universit\`{a} di Salerno,
				Via Giovanni Paolo II,
				$84084$ Fisciano (SA), Italy}}\\
		\small{\textit{$^{(2)}$	Associazione Culturale ``Velia Polis", via Capo di Mezzo
				$84078$, Vallo della Lucania (SA), Italy}}\\
		\small{\textit{$^{(3)}$INFN, Sezione di Napoli, C. U. Monte S. Angelo, Via Cinthia, $80126$ Napoli, Italy}}}

	\title{Microscopic foundations of the Second Law of Thermodynamics within Nonunitary
		Newtonian Gravity}

	\maketitle

\begin{abstract}
	The quest for a microscopic foundation of thermodynamics is addressed within the Nonunitary Newtonian
	Gravity model through the study of a specific closed system, namely a three-dimensional harmonic
	nanocrystal. A numerical calculation of the nanocrystal von Neumann entropy as a function of time is performed,
	showing a sharp monotonic increase, followed by a stabilization at late times.
	This behavior is consistent with the emergence of a micro-canonical ensemble within the initial
	energy levels, signaling, in this way, the establishment of a nonunitary gravity-induced thermal equilibrium.
\end{abstract}

keywords: Gravity; entanglement entropy; thermalization.


\markboth{S. De Filippo, F. Maimone, A. Naddeo, G. Scelza}
{Microscopic foundations of the Second Law of Thermodynamics within Nonunitary
	Newtonian Gravity}

\section{Introduction}	

The two weak points of the final setting of Quantum Mechanics (QM) by von Neumann, the vague notion of a
macroscopic measurement apparatus and the definition of coarse graining entropy,
based, as it is, on the subjective notion of macroscopic observables, can in
principle both be addressed by a nonunitary quantum dynamics. The natural way to get
a nonunitary dynamics is to put the physical system in interaction with an ancillary
system and then tracing out the ancilla to get a mixed state described by a density
matrix, just as it happens with open systems when environment is traced out.
If we want to identify thermodynamic entropy with the entanglement one with an
ancillary or hidden system \cite{unruh,kay,kay1}, we have at least to require thermal
equilibrium between the physical system and the hidden one:
\begin{equation}
\frac{1}{T} = \frac{\partial S}{\partial E_p}=\frac{\partial S}{\partial E_h}=\frac{\partial S}{\partial E_p}\left(\frac{\partial E_h}{\partial E_p} \right)^{-1},
\label{tequilibrium}
\end{equation}
where $E_p$ and $E_h$ denote respectively the energy of the physical and the hidden
system, while the entropy $S$ is one and the same for the two systems, being the
entanglement entropy of a bipartite system.

The above thermal equilibrium condition
leads to the almost inescapable conclusion that every physical system must have as
hidden partner its exact replica and the metasystem (physical plus hidden) state
space must be restricted by a symmetry constraint in the exchange of physical and
hidden degrees of freedom. This constraint eliminates the arbitrariness of
considering as non observable some degrees of freedom, since in any constrained
theory, like Gupta-Bleuler's, the observable algebra is a subalgebra of the original
dynamical one.

Several models were proposed to modify quantum dynamics in order to get localization
and transition to classicality, and some of them tried to establish a link with
gravity \cite{pearle,rimini,weber,diosi}. They all introduced
phenomenological parameters, while for a fundamental theory $G$, $h$ and $c$ should be
enough since these three fundamental constants allow one to make all physical entities
dimensionless. If we limit ourselves to low energy physics, only $G$ and $h$ should appear in an
approximate low energy model and, for dimensional reasons, the only possible
relation between threshold mass $M$ and length $L$ for localization and transition to
classicality is
\begin{equation}
M^3 L = \frac{h^2}{G},
\label{thres1}
\end{equation}
which implies that
\begin{equation}
M = \left(\frac{h^2}{G} \right)^{\frac{3}{10}} \rho^{\frac{1}{10}}.
\label{thres2}
\end{equation}
Here one sees that the dependence on mass density $\rho$ is exceedingly weak: to get
a doubling of $M$, $\rho$ has to get $2^{10}=1024$ times higher. This quasi independence of
the threshold mass on mass density could be an experimental signature of gravitational
localization.

In Refs. \cite{sergio1,sergio2,sergio3,sergio4} a dynamics was defined and analyzed which gives rise to
the same relation as above between threshold mass for localization and mass density. It
introduces a gravitational interaction only between a generic physical system and its replica and
constraints the metastate space by a symmetry requirement as well. In this model no
gravitational interaction was introduced within physical and hidden system as the
author had in his mind the possibility to avoid gravitational-collapse
singularities. But in Refs. \cite{sergio5,sergiofilippo1} it was shown that a general covariant
model, with a sound newtonian limit - differing from the previous one\cite{sergio1,sergio2,sergio3,sergio4} just
for the inclusion of gravity within the physical and the hidden systems, at ordinary
laboratory lengths, and giving about the same localization thresholds - can
eliminate gravitational interaction within physical and hidden systems only for
lengths of the order of Planck length, thus avoiding collapse singularities.

The low-energy limit of this model, a brief account of which is given in the Appendix A, is known as Nonunitary Newtonian Gravity (NNG, from now on), and has
been studied in detail showing (entropic) dynamical self-localization for masses
above the sharp threshold of $10^{11}$ proton masses, with
precise signatures susceptible to future experimental tests \cite{sergiofilippo1,DeFilMaimRob,DeFilMaimEntropy,Nostro}. Recently it has been explicitly shown to be free from causality violation problems \cite{sergio6,nostro3}, at variance with
semiclassical gravity, namely (in the Newtonian limit) Newton-Schroedinger
model. Physically this is achieved by decohering linear superpositions of
macroscopically distinct states, as a consequence of the dynamical mechanism of state reduction naturally embedded in the model.
Another peculiar feature is the evolution of pure states into mixed states even for closed
systems \cite{sergiofilippo1,NostroTwo}, so that within NNG density matrix emerges as the fundamental description of
physical reality. Indeed, to the best of our knowledge, NNG is the first model treating the density matrix
as the fundamental characterization of a closed system state with a non-Markov
evolution even from a pure state. In fact on one hand collapse models do not use
density matrices as fundamental entities but just as associated to a stochastic
evolution, and on the other hand models based on the Lindblad equation give Markov
evolution of a mixed state.

All the above features make the NNG model useful to shed new light on the long standing and still open problem of the foundations of the Second Law of thermodynamics \cite{th2,th3,th6}. Indeed, as pointed out in Ref. \cite{wald}, starting from suitable
initial conditions, only a nonunitary quantum dynamics could allow for a
microscopic derivation of the Second Law of thermodynamics for a closed
system by resorting to von Neumann entropy.

A first step in demonstrating the ability of
NNG to reproduce a gravity-induced relaxation towards thermodynamic
equilibrium even for a closed system has been carried out in Ref. \cite{NostroTwo}, by taking a simple system: two particles in an harmonic trap interacting via an `electrical' delta-like potential and gravitational interaction. Starting from an energy eigenstate, a slow net variation of the von Neumann entropy for the system as a whole has been found, together with a small modulation induced on the relative entanglement entropy of the two particles, while energy expectation remained constant.

The aim of the present paper is to generalize this work to a more complex system, i.e. an harmonic nanocrystal within a cubic geometry. This choice allows us to perform a first step towards the simulation of macroscopic systems where the Second Law of thermodynamics is more relevant. A numerical simulation is carried out by starting from an initial pure physical state with mean energy $E$, drawn uniformly at random by superposing a huge number of energy eigenstates within the energy interval $\Delta E$ around $E$, according to the prescriptions of the Eigenstate Thermalization Hypothesis (ETH) \cite{Deutsch,Rigol}. In this way the behaviour of the von Neumann entropy as a function of time is obtained, which is consistent with the one for a thermalizing system.

The plan of the paper is as follows. In Section II we briefly summarize our recent results on the simple two-particle system. In Section III, which is the core of the paper, we report on the numerical simulation of a harmonic nanocrystal with cubic geometry, pointing out our main results. Finally, in Section IV, we draw some conclusions and outline future
perspectives of this work. A brief general
description of the basic NNG model is given in the Appendix, together with some computational details.

\section{Two-particle simulation: a brief summary of main results}

In this Section we summarize the results of a previous simulation, carried out on a simple system of two interacting particles \cite{NostroTwo}. This has been a first but necessary step in demonstrating the ability of the NNG model to reproduce a gravity-induced relaxation towards thermodynamic
equilibrium even for a closed system \cite{NostroTwo}.

More specifically we consider the two particles in an harmonic trap,
interacting with each other through `electrostatic' and gravitational
interaction, whose `physical' Hamiltonian, in the ordinary
(first-quantization) setting, is
\begin{equation}
H_{ph}\left( \mathbf{x}_{1},\mathbf{x}_{2}\right) =
\sum\limits_{i=1}^{2}\biggl(-\frac{\hbar ^{2}}{2\mu }\Delta _{\mathbf{x}%
	_{i}}+\frac{\mu}{2} \omega ^{2}\mathbf{x}_{i}^{2}\biggr)+\sum\limits_{i<j=1}^{2}\left(\frac{4\pi \hbar ^{2}l_{s}}{%
	\mu }\delta (\mathbf{x}_{i}-\mathbf{x}%
_{j})-\frac{G\mu ^{2}}{\left\vert \mathbf{x}_{i}-\mathbf{x}%
	_{j}\right\vert }\right);  \label{mol1}
\end{equation}
here $\mu $ is the mass of the particles, $\omega $ is the frequency of the
trap and $l_{s}$ is the s-wave scattering length. We are considering a
`dilute' system, such that the electrical interaction can be assumed to have
a contact form with a dominant s-wave scattering channel. We take numerical
parameters that make the electrical interaction at most comparable
with the oscillator's energy, while gravity enters the problem as an higher
order correction. The NNG model, in
the (first-quantization) ordinary setting, is defined by the following general meta-Hamiltonian:
\begin{equation}
H_{TOT}=H_{ph}\left( \mathbf{x}_{1},\mathbf{x}_{2}\right) +H_{hid}\left(
\widetilde{\mathbf{x}}_{1},\widetilde{\mathbf{x}}_{2}\right) +H_{int}\left(
\mathbf{x}_{1},\mathbf{x}_{2};\widetilde{\mathbf{x}}_{1},\widetilde{\mathbf{x%
}}_{2}\right) ,
\end{equation}
with $H_{int}=G\mu^2\sum_{i<j}\biggl(\frac{1}{2\vert \mathbf{x}_i-\mathbf{x}_j\vert}+\frac{1}{%
	2\vert \widetilde{\mathbf{x}}_i-\widetilde{\mathbf{x}}_j\vert}\biggr)-G\mu^2\sum_{i,j}\frac{1}{\vert \mathbf{x}_i-\widetilde{%
		\mathbf{x}}_j\vert}$.

Then the time dependent physical density matrix is computed by tracing out the hidden
degrees of freedom and the corresponding von Neumann entropy is derived as
the entanglement entropy with such hidden degrees of freedom. This is obtained via a numerical simulation, by choosing as initial condition an eigenstate of the physical Hamiltonian. As a result, we find that entropy fluctuations take place, owing to
the (nonunitary part of) gravitational interactions, with the initial pure
state evolving into a mixture \cite{NostroTwo}. The behavior of one- and two-particle von Neumann entropy as a function of time is depicted in Fig. 1 for an initial state $\vert\phi _{2}\rangle$, equal to the eigenstate of the physical energy associated with the 2nd highest energy eigenvalue $E_2$ of the two-particle system under study. The following values of the physical parameters $l_{s}$, $\mu $
and $\omega $ have been chosen: $l_{s}=5.5\cdot 10^{-8}m$, $\mu =1.2\cdot 10^{-24}kg$ and $%
\omega =4\pi \cdot 10^{3}s^{-1}$ (which are compatible with current
experiments with trapped ultracold atoms \cite{exp1} and complex molecules
\cite{exp2}) together with an artificially augmented 'gravitational constant' $%
G=6.67408\times 10^{-6}m^{3}kg^{-1}s^{-2}$ ($10^{5}$ times the real
constant).
\begin{figure}[tbph]
	\centerline{\includegraphics[scale=0.8]{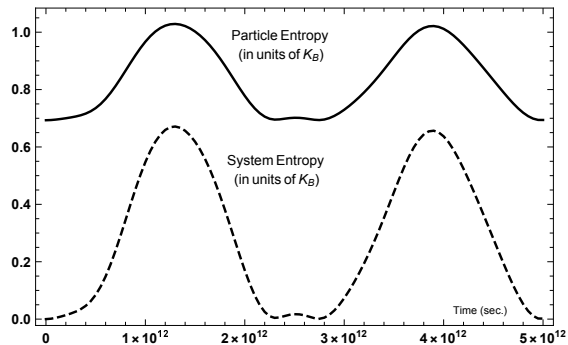}}
	\vspace*{8pt}
	\caption{ Time evolution of one-particle and two-particle entropies (respectively, curve above and below) for the initial state chosen.}
	\label{fig:fig1}
\end{figure}

Due to gravity-induced fluctuations the system
entropy shows a net variation over very long times, at variance with the
case without gravitational interaction with the hidden
system, in which it would have been a constant of motion. At the same
time, single particle entropy, which in the ordinary setting is itself
constant, shows now a similar time modulation.
It was to be expected that for a microscopic system the recurrence times are relatively short, due to the low dimensionality of the Hilbert metaspace corresponding to the microcanonical ensemble; on the contrary the recurrence times of macroscopic systems may be longer than the age of the universe. Actually the recurrence is even more irrelevant since no system can be considered closed for long times.
The expectation of physical energy has been verified to be a constant as well,
meaning that the nonunitary term has no net energy associated with itself,
but is purely fluctuational. In the next Section we extend and generalize these results to a more complex system in order to
further explore how thermodynamics could emerge in the
framework of NNG. In perspective, the authors would like to add new pieces of evidence on the possible role of gravity in the
quantum foundations of thermodynamics in ordinary low-energy physics.

\section{Application to an ideal (harmonic) nanocrystal}

The aim of this Section is to go further in the study of nonunitary gravity-induced
thermalization, thus in the following we switch to a bit more complex model of a
three-dimensional system, that is still computationally tractable. Taking
the simple model of an harmonic nanocrystal, in which we consider as in the
previous case an artificially augmented gravitational constant, turns out to
be a viable choice: indeed the result is a simple and nice formula for thermalization that
can be efficiently simulated numerically.

We consider a cubic crystal of volume $V=L^{3}$. Each phonon propagating
within the crystal is supposed to fill it homogeneously, and its
gravitational mass is given by $\hbar \omega /c^{2}$, where $\omega $ is the
phonon's angular frequency. The total Hamiltonian is given by
\begin{equation}
H_{G}=H_{ph}+H_{hid}+H_{int}
\end{equation}
where
\begin{eqnarray}
H_{ph}&=&\sum_{\mathbf{k}s}\hbar \omega _{\mathbf{k}s}\widehat{n}_{\mathbf{k}s}-\frac{%
	1}{4}C(G,V)\sum_{\mathbf{k}s}\sum_{\mathbf{k}^{\prime }s^{\prime }}\left(
\frac{\hbar \omega _{\mathbf{k}s}}{c^{2}}\right) \left( \frac{\hbar \omega _{%
		\mathbf{k}^{\prime }s^{\prime }}}{c^{2}}\right) \widehat{n}_{\mathbf{ks}}%
\widehat{n}_{\mathbf{k}^{\prime }s^{\prime }}, \\
H_{hid}&=&\sum_{\widetilde{\mathbf{k}}\widetilde{s}}\hbar \omega _{\widetilde{%
		\mathbf{k}}\widetilde{s}}\widehat{n}_{\widetilde{\mathbf{k}}\widetilde{s}}-%
\frac{1}{4}C(G,V)\sum_{\widetilde{\mathbf{k}}\widetilde{s}}\sum_{\widetilde{%
		\mathbf{k}}^{\prime }\widetilde{s}^{\prime }}\left( \frac{\hbar \omega _{%
		\widetilde{\mathbf{k}}\widetilde{s}}}{c^{2}}\right) \left( \frac{\hbar
	\omega _{\widetilde{\mathbf{k}}^{\prime }\widetilde{s}^{\prime }}}{c^{2}}%
\right) \widehat{\widetilde{n}}_{\widetilde{\mathbf{k}}\widetilde{s}}%
\widehat{\widetilde{n}}_{\widetilde{\mathbf{k}}^{\prime }\widetilde{s}%
	^{\prime }} ,\\
H_{int}&=&-\frac{1}{2}C(G,V)\sum_{\mathbf{k}s}\sum_{\widetilde{\mathbf{k}}\widetilde{%
		s}}\left( \frac{\hbar \omega _{\mathbf{k}s}}{c^{2}}\right) \left( \frac{%
	\hbar \omega _{\widetilde{\mathbf{k}}\widetilde{s}}}{c^{2}}\right) \widehat{n%
}_{\mathbf{ks}}\widehat{\widetilde{n}}_{\widetilde{\mathbf{k}}\widetilde{s}%
}\ \ .\ \
\end{eqnarray}
Here $\widehat{n}$ and $\widehat{\widetilde{n}}$ are,
respectively, the physical and hidden phonon number operators. Wave numbers
are given by
\begin{equation*}
k_{i}=\frac{2\pi n_{i}}{L},\ \ \text{\ with}\ \ n_{i}=0,\pm 1,\pm 2,...\ \ \
\text{and}\ \ -\frac{\pi }{a}<k_{i}\leq \frac{\pi }{a}\ \ \ \ \ \text{(first
	Brillouin zone),}
\end{equation*}
while the gravitational factor $C(G,V)$, linearly depending on $G$, is
calculated in Appendix B.

We assume a simple dispersion relation, corresponding to a simple
cubic crystal structure in which only the first $6$ neighbors interaction is
taken into account:
\begin{equation}
\omega _{\mathbf{k}s}=\sqrt{\frac{4K}{m}}\left\vert \sin \left( \frac{ak_{s}%
}{2}\right) \right\vert ,\ \ \ \ \ \ \ \ \ \ \ \ \ s=1,2,3
\end{equation}
where $m$ is the atomic mass, $K$ is the elastic constant and $a$ is the
lattice constant.

Indicating by $\left\vert \mathbf{n}\right\rangle =\left\vert n_{\mathbf{k}%
	_{1}s_{1}}n_{\mathbf{k}_{2}s_{2}}...\right\rangle $ the state number in the
physical Fock space and by $\left\vert \widetilde{\mathbf{n}}\right\rangle
=\left\vert \widetilde{n}_{\mathbf{k}_{1}s_{1}}\widetilde{n}_{\mathbf{k}%
	_{2}s_{2}}...\right\rangle $ the state number in the hidden Fock space, we
note that two generic state numbers $\left\vert \mathbf{n}^{i}\right\rangle $
and $\left\vert \widetilde{\mathbf{n}}^{j}\right\rangle $ are respectively
eigenstates of $H_{ph}$ and $H_{hid}$. This follows from the simple
observation that these Hamiltonians depend only on their respective number
operators. Let's call $E_{0,i}$ and $E_{0,j}$ their respective eigenvalues.
Now, the product $\left\vert \mathbf{n}^{i}\right\rangle \otimes \left\vert
\widetilde{\mathbf{n}}^{j}\right\rangle $ is an eigenstate of the total
Hamiltonian $H_{G}$ with eigenvalue $E_{0,i}+$ $E_{0,j}+E_{Int,i,j}$. As for
a really macroscopic (or mesoscopic) body, thermodynamic variables like energy can be defined only at a macroscopic level \cite{th2}. In particular, a thermodynamic state with
internal energy $E$ amounts, at the microscopic level, to specifying energy
within an uncertainty $\Delta E$, which includes a huge number of energy
levels of the body. For this reason, assuming an initial pure physical state
of the form
\begin{equation}
\left\vert \psi \left( 0\right) \right\rangle =\sum_{i}\gamma _{i}\left\vert
\mathbf{n}^{i}\right\rangle ,  \label{RandomState}
\end{equation}
where $\left\vert \mathbf{n}^{i}\right\rangle $ are the (physical)
energy eigenstates with eigenvalues $E_{0,i}$ close to $E$ (within $\Delta E$%
), the corresponding meta-state is:
\begin{equation}
\Vert \Psi \left( 0\right) \rangle \rangle =\left\vert \psi \left( 0\right)
\right\rangle \otimes \left\vert \widetilde{\psi }\left( 0\right)
\right\rangle .
\end{equation}
The state (\ref{RandomState}) is intended to be drawn uniformly at random from
the high dimensional subspace corresponding to the energy interval
considered. This is reminiscent of the notion of \textit{tipicality} introduced
in the context of the Eigenstate Thermalization Hypothesis (ETH) \cite{Rigol}%
, where this concept is more precisely stated by saying that state vector
above is distributed according to the Haar measure over the considered
subspace \cite{Gogolin}. To the purpose of our numerical implementation, the
simple algorithm described in Ref. \cite{Maziero} is used.

At time $t$, assuming a complete isolation of the body, we get
\begin{equation}
\Vert \Psi \left( t\right) \rangle \rangle =\sum\limits_{i,j}\gamma
_{i}\gamma _{j}e^{-\left( i/\hbar \right) \left[ E_{0,i}+E_{0,j}+E_{Int,i,j}%
	\right] \ t}\left\vert \mathbf{n}^{i}\right\rangle \otimes \left\vert \widetilde{\mathbf{n}}%
^{j}\right\rangle \equiv \Gamma _{i,j}\left( t\right) \left\vert
\mathbf{n}^{i}\right\rangle \otimes \left\vert \widetilde{\mathbf{n}}^{j}\right\rangle .
\end{equation}
The physical state $\rho _{ph}$ is then given by
\begin{equation}
\rho _{ph}\left( t\right) =\sum\limits_{i,i^{\prime }}f_{i,i^{\prime
}}\left( t\right) \left\vert \mathbf{n}^{i}\right\rangle \left\langle
\mathbf{n}^{i^{\prime }}\right\vert ,
\end{equation}
with
\begin{eqnarray}
f_{i,i^{\prime }}\left( t\right) &=&\sum\limits_{j}\Gamma _{i,j}^{\ast }\left(
t\right) \Gamma _{i^{\prime },j}\left( t\right)=\gamma _{i}^{\ast }\gamma _{i^{\prime }}e^{-(i/\hbar )\left[ E_{0,i^{\prime
	}}-E_{0,i}\right] t} \sum\limits_{j}\left\vert \gamma _{j}\right\vert^{2} \nonumber \\
&\times & \exp \left\{  \frac{i}{2\hbar}  C(G,V)\sum\limits_{\mathbf{k}s,%
	\mathbf{k}^{\prime }s^{\prime }}\left( \frac{\hbar \omega _{\mathbf{k}s}}{%
	c^{2}}\right) \left( \frac{\hbar \omega _{\mathbf{k}^{\prime }s^{\prime }}}{%
	c^{2}}\right) \left[ n_{\mathbf{k}s}^{i^{\prime }}-n_{\mathbf{k}s^{\prime
}}^{i}\right] n_{\mathbf{k}^{\prime }s^{\prime }}^{j}t\right\} .
\label{coeffi}
\end{eqnarray}
This last formula is our central result. In fact the term within the
square brackets is the one responsible for the rapid phase cancelation and
diagonalization of $\rho _{ph}\left( t\right) $ in the energy basis, as we
show in the numerical simulation that follows. Incidentally, we note the
strict resemblance of Eq. (\ref{coeffi}) with Eq. (51) of Ref. \cite{sergiofilippo1}%
, expressing the phases cancelation leading to the dynamical
self-localization of a lump. This fact reflects the deep connection between the quantum measurement problem and the law
of entropy increase, as pointed out in Ref. \cite{th2}.

In order to perform a simple and viable numerical simulation of the time
evolution of the system through the explicit computation of all the terms in Eq. (%
\ref{coeffi}), we consider a nanocrystal of $10^{3}$atoms, with
the following values for the parameters: $m=3.48\times 10^{-25}kg$ (i.e. the mass of $^{210}Po$, the only chemical
element presenting a simple cubic crystal structure), $a=335\ pm$ ($L=9a$) and
$K=23.091\ N/m$. We put a huge factor $F=10^{46}$ in front of $G$ in
order to simulate the effect of gravity in a really macroscopic system
(otherwise the characteristic time of gravitational thermalization for a
system of only $10^{3}$ atoms would be much greater than the age of the
Universe!). Besides we choose $E=1.89\ \times 10^{6}\ \hbar \sqrt{K/m}\ $and
$\Delta E$ $=0.0024\ E$.

\begin{figure}[tbph]
	\centerline{\includegraphics[scale=0.4]{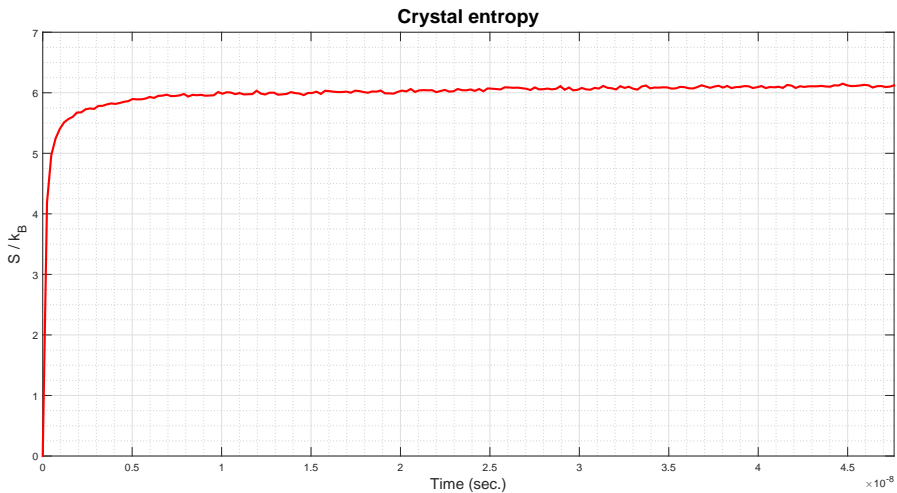}}
	\vspace*{8pt}
	\caption{Entropy as a function of time for an initial superposition of energy eigenstates within
		the interval $\Delta E$, showing a monotonic increase and
		stabilization at late times, as expected for a thermalizing system.}
	\label{figure2}
\end{figure}

The numerical calculation of von Neumann entropy amounts to the repeated
diagonalization, on a discretized time axis, of the numerical matrix $%
f_{i,i^{\prime }}$. Denoting with $\lambda _{j}$ the eigenvalues of this latter
matrix at a given time, von Neumann entropy is readily computed as $%
S/k_{B}=-\sum_{j}\lambda _{j}\ln \lambda _{j}$. Its time behaviour is
shown in Fig. \ref{figure2}. We can see that the gravitational term at
work reproduces correctly the expected behaviour of a thermalizing system.
It is expected that the final value of entropy is the maximum value
attainable at the given internal energy $E$, provided that the state $\rho _{ph}$
contains all the available energy eigenstates (given the supposed \textit{%
	typicality} of the initial state). Since the off-diagonal terms of $%
\rho _{phys}$ quickly die out in the basis of the physical energy,
consistency with the micro-canonical ensemble, and then with Thermodynamics,
is ensured. To be more precise, as it can be immediately
seen from Eq. (\ref{coeffi}), coherences still survive within the degenerate subspaces
of energy associated, in the case under study, to a permutation
symmetry of the branches $\alpha $. This amounts to a erroneous factor $3!$ in
the counting of states, which is practically irrelevant in the computation
of entropy due to the huge number of states involved. The characteristic time
for entropy stabilization depends of course on the factor $F$ multiplying $G$%
, that we have inserted to mimic the effect of a really macroscopic crystal.
Reducing $F$ amounts to an increase of this time, being the two parameters
inversely proportional, thanks to the time-energy uncertainty relation.

Up to now we have tacitly assumed in our model that the physical and hidden
crystals are perfectly superimposed on each other. This circumstance holds
true only if a strict CM\ self-localization is verified, i.e. if our crystal
mass is well above the gravitational localization threshold. In the case
under study, the big factor $F$ multiplying $G$ has to be taken into
account. Localization length can be estimated by using $\Lambda \sim \left(
\hbar \sqrt{V\ /\ FGM^{3}}\right) ^{1/2}\sim 10^{-15}m<<L\sim 10^{-9}m$.
This latter condition ensures the physical consistency of our calculation.

Finally, let us stress that the present model should be considered
just as a toy-model of a real crystal. In fact, in a real crystal the
anharmonic corrections play an important role in subsystems' thermalization,
and are of lower order with respect to (nonunitary) gravity, this latter
being qualitatively different because of its nonunitary nature. Indeed it is just the nonunitarity which gives rise
to the possibility of a net entropy growth for the system as a whole, so
allowing for a microscopic derivation of the Second Law of
thermodynamics. While, of course, the general model need to be tested
against properly designed future experiments, anyway we can say that it is
the first self-consistent low-energy gravity model implying in a natural way the
emergence of Thermodynamics even in a closed system.
Incidentally, within the framework of ETH \cite{Deutsch,Rigol}, a physical entropy can only be
introduced by previously applying an ad-hoc \textit{de-phasing map} to the
state of the system.

\section{Conclusions and perspectives}
In this work the ability of the NNG model to produce gravity induced thermalization in a closed quantum many-body system has been investigated by studying a specific three-dimensional harmonic nanocrystal model. A numerical simulation compatible with the ETH prescriptions for the choice of the initial state \cite{Deutsch,Rigol} has been carried out, and the time evolution of the system has been calculated. The result shows a monotonic increase of the von Neumann entropy, followed by a stabilization at late times. This behaviour is consistent with that of thermalizing system.

As to the future perspective of this work, the
numerical simulation of a more realistic (mesoscopic) crystal is currently under study, in which on
the one hand the real value of $G$ is used, and on the other hand the
anharmonic terms are explicitly included. These terms are supposed to play
an important role to speed up thermalization by letting the phonons
interact. They are just the analogue of the electromagnetic interaction acting in the
two-particles system discussed in Ref. \cite{NostroTwo} and recalled in Section 2. It is expected, within this more realistic
setting, that even starting from a sharp energy level of the whole crystal,
a micro-canonical ensemble would emerge, putting eventually the mechanism of
nonunitary gravity-induced thermalization on more robust grounds. If this
happens, \textit{typicality} of the initial state need not to be assumed
\textit{a priori}.



\appendix
\section{NNG model: a brief account}

In this Appendix we summarize the
main features of the NNG model, by focusing on its simplest form, suitably oriented to the problem under study in the main text \cite{sergio1,sergio2,sergio3,sergio5,sergiofilippo1}.

Let $\mathcal{H}_{0}[\psi ^{\dagger },\psi ]$ be the non-relativistic
`physical' Hamiltonian of a finite number of particle species, like
electrons, nuclei, ions, atoms and/or molecules, where $\psi ^{\dagger
},\psi $ denote the whole set $\psi _{j}^{\dagger }(\mathbf{x}),\psi _{j}(%
\mathbf{x})$ of creation-annihilation operators, \textit{i.e.} one couple
per particle species and spin component. $\mathcal{H}_{0}[\psi ^{\dagger
},\psi ]$ includes the usual electromagnetic interactions accounted for in
atomic, molecular and condensed-matter physics.

Denoting by $H_{ph}$ the `physical' energy operator including also the usual
Newtonian interaction,
\begin{equation}
\mathcal{H}_{ph}[\psi ^{\dagger },\psi ]=\mathcal{H}_{0}[\psi ^{\dagger
},\psi ]-\frac{G}{2}\sum_{j,k}m_{j}m_{k}\int d\mathbf{x}d\mathbf{y}\frac{%
	:\psi _{j}^{\dagger }(\mathbf{x})\psi _{j}(\mathbf{x})\psi _{k}^{\dagger }(%
	\mathbf{y})\psi _{k}(\mathbf{y}):}{|\mathbf{x-y}|},
\end{equation}
to incorporate that part of gravitational interactions responsible for
non-unitarity one has to introduce complementary creation-annihilation
operators $\widetilde{\psi }_{j}^{\dagger }(\mathbf{x}),\widetilde{\psi }%
_{j}(\mathbf{x})$ and the overall (meta-)Hamiltonian:
\begin{eqnarray}
\mathcal{H}_{TOT}& =&\mathcal{H}_{ph}[\psi ^{\dagger },\psi ]+\mathcal{H}%
_{hid}[\tilde{\psi}^{\dagger },\tilde{\psi}] \nonumber \\
& -&\frac{G}{4}\sum_{j,k}m_{j}m_{k}\int d\mathbf{x}d\mathbf{y}\left[ \frac{%
	2\psi _{j}^{\dagger }(\mathbf{x})\psi _{j}(\mathbf{x})\tilde{\psi}%
	_{k}^{\dagger }(\mathbf{y})\tilde{\psi}_{k}(\mathbf{y})}{|\mathbf{x-y}|}%
\right] \nonumber \\
& +& \frac{G}{4}\sum_{j,k}m_{j}m_{k}\int d\mathbf{x}d\mathbf{y}\left[ \frac{%
	:\psi _{j}^{\dagger }(\mathbf{x})\psi _{j}(\mathbf{x})\psi _{k}^{\dagger }(%
	\mathbf{y})\psi _{k}(\mathbf{y}):}{|\mathbf{x-y}|}\right] \nonumber \\
& +&\frac{G}{4}\sum_{j,k}m_{j}m_{k}\int d\mathbf{x}d\mathbf{y}\left[ \frac{%
	:\tilde{\psi}_{j}^{\dagger }(\mathbf{x})%
	\tilde{\psi}_{j}(\mathbf{x})\tilde{\psi}_{k}^{\dagger }(\mathbf{y})\tilde{%
		\psi}_{k}(\mathbf{y}):}{|\mathbf{x-y}|}\right].
\label{ge}
\end{eqnarray}
The above operators act on the product $F_{\psi }\otimes F_{\widetilde{\psi }%
}$ of Fock spaces of the $\psi $ and $\widetilde{\psi }$ operators, where $%
m_{i}$ is the mass of the $i$-th particle species and $G$ is the
gravitational constant. The $\widetilde{\psi }$ operators obey the same
statistics as the corresponding operators $\psi $, while $[\psi ,\widetilde{
	\psi }]_{-}=[\psi ,\widetilde{\psi }^{\dagger }]_{-}=0$.

The meta-particle state space $S$ is the subspace of $F_{\psi }\otimes F_{
	\widetilde{\psi }}$, including the meta-states obtained from the vacuum $%
\left\vert \left\vert 0\right\rangle \right\rangle =\left\vert
0\right\rangle _{\psi }\otimes \left\vert 0\right\rangle _{\widetilde{\psi }%
} $ by applying operators built in terms of the products $\psi _{j}^{\dagger
}( \mathbf{x})\widetilde{\psi }_{j}^{\dagger }(\mathbf{y})$ and symmetrical
with respect to the interchange $\psi ^{\dagger }\leftrightarrow \widetilde{
	\psi }^{\dagger }$; as a consequence they have the same number of $\psi $
(physical) and $\widetilde{\psi }$ (hidden) meta-particles of each species.
Since constrained meta-states cannot distinguish between physical and hidden
operators, the observable algebra is identified with the physical operator
algebra. In view of this, expectation values can be evaluated by
preliminarily tracing out the $\widetilde{\psi }$ operators. In particular,
the most general meta-state corresponding to one particle states is
represented by
\begin{equation}
\left\vert \left\vert f\right\rangle \right\rangle =\int d\mathbf{x}\int d%
\mathbf{y}\,f(\mathbf{x},\mathbf{y})\psi _{j}^{\dagger }(\mathbf{x})%
\widetilde{\psi }_{j}^{\dagger }(\mathbf{y})\left\vert \left\vert
0\right\rangle \right\rangle ,  \label{f}
\end{equation}%
with
\begin{equation*}
f(\mathbf{x},\mathbf{y})=f(\mathbf{y},\mathbf{x}).
\end{equation*}%
This is a consistent definition since $\mathcal{H}_{TOT}$\ generates a group
of (unitary) endomorphisms of $S$.

Note that $\mathcal{H}_{TOT}$ and $\mathcal{H}_{ph}[\psi ^{\dagger },\psi ]+%
\mathcal{H}_{hid}[\tilde{\psi}^{\dagger },\tilde{\psi}]$ differ only due to
correlations. In fact, because of the state constraint, the hidden degrees
of freedom show the same average energy of the observed ones, while the two
last sums in $\mathcal{H}_{TOT}$ have approximately equal expectation values
and fluctuate around the classical gravitational energy. These fluctuations, though irrelevant
on a macroscopic scale, are precisely what can lead to thermodynamic
equilibrium in a closed system if thermodynamic entropy is identified with
von Neumann entropy \cite{kay1}.

\section{Calculation of $C(G,V)$}

Let's calculate the gravitational interaction between two phonons, supposed
to fill homogeneously the crystal of cubic form. Their mass are expressed by
\begin{equation}
m_{i}=\frac{\hbar \omega _{\mathbf{k}_{i}}}{c^{2}},\ \ \ \ \ \ \ \ \ \
i=1,2\
\end{equation}
while their gravitational interaction energy is
\begin{equation}
U_{G}=-m_{1}m_{2}C(G,V)=-m_{1}m_{2}\ G\ V^{-1/3}\ \Im ,
\end{equation}
where
\begin{equation}
\Im =\int\limits_{0}^{1}d\xi \int\limits_{0}^{1}d\eta
\int\limits_{0}^{1}d\zeta \int\limits_{0}^{1}d\xi ^{\prime
}\int\limits_{0}^{1}d\eta ^{\prime }\int\limits_{0}^{1}d\zeta ^{\prime }%
\frac{1}{\sqrt{\left( \xi -\xi ^{\prime }\right) ^{2}+\left( \eta -\eta
		^{\prime }\right) ^{2}+\left( \zeta -\zeta ^{\prime }\right) ^{2}}}\simeq
1.87.
\end{equation}
\bigskip
This value has been obtained by means of two different numerical methods
with the software @Mathematica.

\end{document}